# HUMAN COMPUTER SYMBIOSIS

*Eluyefa Olanrewaju Andrew*
**IT Project Management**
**Aston University**
**Birmingham**
**eluyefao@aston.ac.uk**

## Abstract

Human Computer Symbiosis is similar to Human Computer Interaction in the sense that it is about how humans and computer interact with each other. For this interaction to be made there needs to be a symbiotic relationship between man and computer. Man can interact with computer in many ways, either just by typing with the keyboard or surfing the web. The cyber-physical-socio space is an important aspect to be looked into when referring to the interaction between man and computer. This paper investigates various aspects related to human computer symbiosis. Alongside the aspects related to the topic, this paper would also look into the limitations of Human Computer Symbiosis and evaluate some previously proposed solutions.

## I. INTRODUCTION

Symbiosis is the mutual benefit or relationship established with long term interaction between species. Symbiosis is of two main types which are; strong symbiosis and weak symbiosis. Strong symbiosis influences the survival of species while weak symbiosis is beneficial but is not totally necessary for the survival of species.

Human computer symbiosis is important as it aims to bring the computing machine effectively into the productive parts of technical problems [13]. The other aim is closely related. This aim is to bring computing machines effectively into processes of thinking that must go on in real time; time that moves too fast to permit using computers in conventional ways [13].

This paper would go in details about the importance of human computer symbiosis, what it aimed to achieve from the onset, identify the progress made so far and if the original aims have been achieved in modern times.

The paper would further look into several aspects like the cyber-physical-socio spaces, symbiosis interaction, the limitations of human computer symbiosis, human computer symbiosis as an enterprise information system and a comparison of the human computer symbiosis with other information systems.

1. **ORIGIN OF HUMAN COMPUTER SYMBIOSIS**

The first mention of human computer symbiosis was made in 1960 by Licklider. His paper titled "*Man Computer Symbiosis*" was made to present the concept of the topic and to foster its development by analyzing some problems of interaction between men and computing machines, which would call to attention to applicable principles of man machine engineering and pointing out a few questions to which research answers were needed.

Licklider hoped to achieve that in the years to come, the human brain and computing machines would be linked tightly together and that the final outcome or partnership will be that the reasoning of the human brain would be like no human brain has ever reasoned, and process data in ways not come across by the information handling machines of that generation [13] [30].

Symbiosis is all about having an interaction between two or more species. In biological terms, it is defined as any living arrangements between members of two different species including mutualism, commensalism, and parasitism. Also any associations between two species populations that live together are symbiotic whether beneficial, harmful or have no effect on one another [3].

The need for interaction between man and computer is very essential, as man on his own could be noisy and flexible, self-programming, have parallel and simultaneously active nervous channels and finally have flexible and rich languages, while computers on its own could be fast and accurate, perform one or more elementary operations at a time, pre-programming and finally have rigid and simple languages.

The essential aim of the symbiosis is to be able to integrate the positive characteristics of both man and computer. This is because computer can do things that are difficult for man, and man can also do many things that are impossible for computer. Hence, having a symbiotic relationship would make for a great value, and provide beneficial outcome, but that can be achieved if the integration of the positive characteristics of both man and computer is successful.

In order for human computer symbiosis to be carried out, there has to be several functions that would be separated by man and computer. The human functions are to set the goals, motivation, hypothesis, and questions. Also, to carry out processes, models and remember past models.

Finally, the function of man is to define the criteria to be used, and handle low probability situations.

The functions to be carried out by computers are to test the model against data, carry out routine and clerical operations, make statistical inference and mathematical calculations and finally, to make diagnosis, pattern matching and relevancy recognition[13] [30].

The common function of man and computer is the decision made based on the data provided. As mentioned earlier, this paper will look at the intended aim of human computer symbiosis made by Licklider (1960) and the current state of the human computer symbiosis in the modern times. Below is a table showing the perception of Licklider and the current version of the human computer symbiosis [13] [30].

| LICKLIDER'S PERCEPTION | CURRENT VERSION |
|---|---|
| Speed Matching<br>• Think Centre – online digital library + workflow | Speed Matching<br>• Many (network)-to-Many (society) |
| Memory hardware requirement<br>• Organize data and texts<br>• Indelible memory + Published memory<br>• As fast as processing unit (speed) | Memory hardware requirement<br>• Organize social media<br>• Cyberspace: CD, Web, Cloud<br>• Speed (still a challenge) |
| Memory Organization | Memory Organization |
| Language | Language (still a challenge) |
| Input Output equipment<br>• Desk surface, wall display<br>• Speech production and recognition | Input Output equipment<br>• Multi-media<br>• Speech, gesture recognition<br>• 3D print |

## 2. SYMBIOTIC INTERACTION

Symbiotic interaction detects interactive computerized systems that can create further steps with respect to user-centered design paradigms [1]. It can be achieved through the combination of computation, sensing technology, and interaction design to gain deep perception, awareness, and understanding between humans and computers. There are significant aspects to execute in symbiotic interaction. They are as follows;

- Transparency, as the property of the computing that would be held responsible.
- Reciprocity and collaborative use of resources for both humans and computers.
- Symbiotic relationships; these are also distinguished by goals and agency independence of humans and computers.

Aside from symbiotic interaction, there are other closely related frameworks which symbiotic interaction has been compared with in terms of system properties. These system properties in question are as follows; understanding context, model cognitive state, detect emotion, affect behavior, induce experience, affect emotion, manipulate and accountable. The diagram below shows how symbiotic interaction has a comprehensive relationship with the system properties [1].

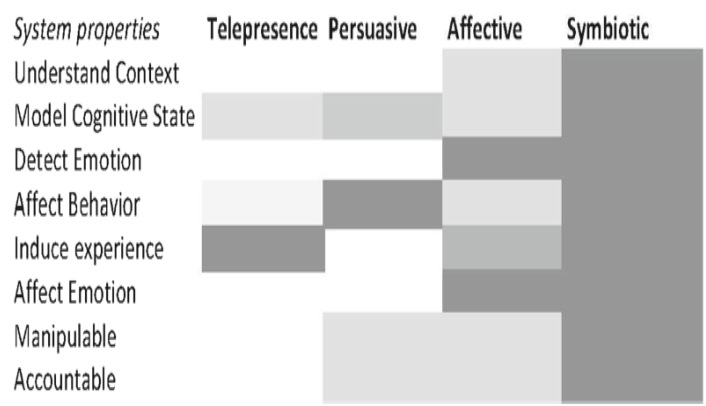

From the above diagram, all the frameworks have at least one relationship with the system properties. In the case of the Affective computing framework, it has a relationship with all the properties but not as comprehensive with all as the Symbiotic framework has.

Telepresence framework consists of subjective experience due to being in a mediated environment; such environment is mainly supported through means of digital resources [18]. It is also a notion used in human computer interaction and media studies. Telepresence channels the intimacy of the relation between humans and the tools we use. This framework describes a core aspect about symbiosis which is the interdependence between the units in a symbiotic relation, in this case, human and computer.

In terms of Persuasive Technologies framework, there are several applications which can aid people in having a change of habit. These applications can direct users towards specific behavioral options in a variety of areas such as sustainable choices and pro-environmental awareness, health and wellness, safety etc. [5].

In other words, Persuasive Technologies framework is all about the influencing of individuals to adopt a certain behavioral pattern through the aid of technology [1].

Affective Computing framework deals with computing that relates to, arises from, or intentionally influences emotions [17]. In general, this leads to research on affective input and affective output of a computing system. Another research involved in affective computing is the advancement of embodied agents that have the ability of expressing emotions, as a vital step is to reenact affective and conversational behaviors [2].

Affective computing is relevant to symbiotic interaction as it is focused on detecting or affecting emotions, which in turn can deepen the symbiotic relationship between humans and machines [17].

3. **CYBER-PHYSICAL-SOCIO SPACES**

This paper would look into enterprise information systems and these enterprises run in a cyber-physical society. The cyber-physical society can also be called the cyber-physical-socio space, known to be a combination of the physical space, the cyber space and the socio space.

A cyber-physical society is a complex space that generates and evolves diverse subspaces to contain various individuals interacting with, reflecting or influencing each other directly or through spaces. The characteristics of the individuals are that they are healthy, they co-evolve harmoniously, provide on-demand information, knowledge and services, they transform from one form into another, they interact with each other through various links, they self-organize according to socio value chains, and they expand according to overall capacity and various flows [28] [30].

The links which the individuals use to interact are the hyperlink, semantic link and complex link.

Hyperlink uses keywords and cyberspace addresses. Semantic link uses class hierarchy, rules and nodes as class. Complex link uses multiple space modelling, physical rules, physiological rules, psychological rules, socio rules, various flows and node as service [30].

*3.1.* **Hyperlink**

Engelbart (1963) proposed a conceptual framework for the augmentation of the intellect to man. It was made known that a certain amount of reference links could be developed between statements within and between files [7].

Hyperlink has the following characteristics:
- The construction relies on programmers
- The labels given by the programmers are not enough to express the relations between texts.
- Hyperlink denotes a reference relation and enables reference texts to be available for readers immediately.
- No restriction when inserting a hyperlink to a hypertext.
- It does not support reasoning. In other words, no hyperlinks can be derived from any existing hyperlinks.
- It does not concern linking rules.
- It does not concern the effect of adding or removing hyperlink [28].

### 3.2. Semantic Link

In order to have rich socio relations and reflect the semantics of real socio networks, a set of semantics links such as implication, subtype, cause-effect and reference as well as a set of reasoning rules were suggested [22].

The assignment of semantic rules to links enables some implicit links to be derived and some semantics to emerge with network motion [26]. A basic semantic space for regulating the semantics of a Semantic Link Network consists of a classification hierarchy of concepts and a set of linking rules [26].

The pioneer gimmick of the semantic link networks are the following:

- To give support to intelligent applications by means of assigning semantic indicators and rules to links and approving relational reasoning, analogical reasoning, inductive reasoning and complex reasoning.
- To analyze the laws of semantic linking.
- To provide with a light-weight semantic networking approach to peer-to-peer knowledge sharing [25].

There are four levels of pursuits of semantic linking [28].

i. The extension of the hyperlink network to aid some elementary intelligence such as guided browsing, query constructive relations, question answering and explanation by the introduction of a semantic space and reasoning mechanisms such as relational reasoning, analogical reasoning, inductive reasoning, and complex reasoning.

ii. The establishment of a self-organized semantic networking model to aid basic socio intelligence with some highly regarded features such as self-organized peer-to-peer linking and high-performance routing, semantic communities and developing principles, common and effect-aware linking, and link recommendation.

iii. The realization of semantic lens by assimilating with the multi-dimensional classification space [27].

iv. The exploration of the general linking methodology. This deals with the new philosophical thinking, interactive computing model, and nature of the complex linking such as dynamicity, symmetry and rules of various flows, the methods for the improvement of various closed loops, and the methods for the coordination of spaces, controlling and predicting evolution.

### 3.3. Complex Link

Here, we would look at four important complex links. These links are Mental-Cyber-Mental Space (MCM), Mental-Cyber-Physical Space (MCP), Mental-Cyber-Socio Space (MCS) and Mental-Artifact-Cyber Space (MAC) [28].

The Mental Space-Cyber Space-Mental Space is the interaction between mental spaces through the cyber space. This complex link can be said to be a service that can facilitate interaction in life-time. This link can extract the Sematic Link Network, discussed earlier, from interactions and make necessary abstraction and evolution during interaction based on the design of the mental space and the translation between languages under different cultures. MCM can recover previous interaction topics, confirm potential partners, and affect reasoning while interacting.

The Mental Space-Cyber Space-Physical Space facilitates the interaction between the mental space and the physical space. This link can be referred to as a service that can obtain the characteristics of the physical objects, and enable the mental space to reflect the real-time status of the physical objects. Sensors are able to emulate some surface features of physical objects. Those well suited to know where, how many, and what types of sensors or actuators needed and how they are arranged to reflect the real-time status of an artifact are the designers.

The Mental Space-Cyber Space- Socio Space eases the interaction between the mental space and the socio space through the cyber space. This link is advertised as a service that can shape the mental space, determine socio behaviors and events, and be responsible for services according to the semantic image emerging in the mental space and socio rules.

The Mental Space-Artifact Space-Cyber Space eases the interaction between the mental space and the cyber space through the artifact space [28]. This complex link allows humans to comprehend and influence the cyber space through adaptable artifacts such as robots in the artifact space. Such robots that are linked to the cyber space could also form cyber semantic images to reflect the physical space and themselves and to share with others [28].

Complex link network have the following characteristics:

- Diversity: To transmit diverse resources such as information, knowledge and energy [16]. Complex link can link various individuals in diverse spaces to aid creation and well-being.
- Real-time influence through spaces
- Cyber-physical-socio context. Complex links can be used as the situation of interactions. It can adapt richer semantics than previous forms of context in the cyber space.
- Cyber-physical-socio service. These links connect various services in different spaces to administer cyber-physical-socio services for individuals. An example is that complex links can link services of recommending universities to the top degree, to the best professors, and to the students.

The cyber space is of two types; one type that allows users to read information in the cyber space like the web and the other type that allows users to read and write information in the cyber space like the web 2.0. Both types of cyber spaces rely on humans to add information to the cyber space for sharing with others [28].

Cyber-physical space is an extension of the cyber space to the physical space. This is done through the use of various sensors. Some unique information in the physical space can be automatically sensed, stored and transmitted through the cyber space. A good example of the cyber-physical space would be the Internet of Things. By giving physical objects a unique identifier, the information from the physical object would be available on the internet hence the extension of the cyber and physical spaces.

This leads to the cyber-physical human space, which extends the cyber-physical space. Here, the behavior of a user(s) can be retrieved and fed back to the cyber space for analysis. These analyses are based on human patterns in how one carries out a certain workout routine or a daily routine in terms of living. Also the user(s) can control the motor used for moving or controlling a mechanism to behave in the physical space through the cyber space [28]. Based on the above explanation of the various spaces, cyber-physical-socio space can be better understood by stating that not only individual's behaviors can be fed back to the cyber space but also the socio interactions as well can be fed back for further processing. The sensors are limited in the ability to retrieve all information in the physical space, so users still need to directly collect the important information in the physical space and afterwards input them into the cyber space after further analysis [28]. Furthermore, the users can change physical objects in the physical space, which could be used as feedback in the cyber space to reflect the real time situation.

Hai Zhuge extended the notion of human computer symbiosis to the notion of human-machine-nature symbiosis as one of the basic concepts of developing the future cyber-physical society [30] [31].

4. **LIMITATIONS OF HUMAN COMPUTER SYMBIOSIS**

This paper has above made mention of the importance of the human computer symbiosis and has also stated its aims. However, the human computer symbiosis has its limitations. This section is going to discuss such limitations and give proposed solutions as to how to overcome such limitations if possible.

I mentioned earlier in this paper Licklider's perception for human computer symbiosis, and in a tabular form showed what is currently available as to what was perceived [13]. It is clear that not all the goals set by Licklider have been achieved till date.

There are two main aspects that have not been achieved hence making them the limitations to human computer symbiosis. The two aspects are speed and language. These posed difficulties when Licklider was dealing with the symbiosis and till date still poses some difficulties.

Foster (2007) proposed four key areas which through the use of improved tools, one could further advance the goals set for the enhancement of the human intellect. These four areas are: *service oriented science, provenance, knowledge communities, and automation of problem-solving protocols* [12].

These fours can be looked upon as potential solutions to the limitations of human computer symbiosis as those four areas would aid in the increase of human intellect, thereby advancing the human aspect of human computer symbiosis.

## 4.1. Service Oriented Science

Service Oriented Science (SOS) is systems that are structured to perform complex computational tasks in terms of communicating services. The SOS is important for handling rapid growth of the volume of data and the increase in complexity of computing and research. SOS methods bring to reality the decomposing and distribution of responsibility for complex tasks in order for the participation of individuals in the construction of a possible solution.

In order for the successful realization of SOS, the following would be needed [10]:

- Resources: these include data, software, and sensors. These help to motivate individuals to develop and operate services that grant access to those resources.
- Supporting Software and Services: these entitles the clients to discover, and determine which services would satisfies their requirements, and also help the clients comprehend the outcome gotten back from the services.
- Hardware Infrastructure: these alongside operational support and policies grant services to be carried out in an appropriate, reliable, and secure manner. In return that allows users to gain access to services efficiently over local area networks and wide area networks.
- Education and Training: the technical expertise required for constructing, operating, and use services are taught in order to develop a community of developers, operators and users.

For this to be a success, each area would depend on both technological and sociological issues to also be successful [10].

## 4.2. Provenance

This is a term that is often used when a source of goods i.e. a computer hardware, is ascertain whether or not it is a genuine product or a counterfeit.

Scientifically speaking, for progress to be made, it depends on a researcher's ability to capitalize on the results gotten from another researcher. These results gotten from another researcher can be classified as useful if other researchers can ascertain how credible they are, and also convince those to whom they propose their ideas to, based on the results, that the originating result is from a credible source [12].

One way to go about this issue would be to lie emphasizes on the reputation of the author which would be used as a basis for evaluation. In general, people tend to trust research papers from an author to whom they have read a previous paper(s) or one that is known as a renowned researcher is his or her field.

Apart from the author's reputation, another way to ascertain the authenticity of the result is the methods used to get such results. The details on the methods used to obtain the results have to be accurate in order for the new researcher to get an insight as to how such results were obtained and also help for further studies [12].

### 4.3. Knowledge Communities

Communities can be said to be a place amongst places where research is carried out. A research laboratory where scientists carry out various experiments to prove a theory can be said to be a community. These communities are formed and operated through the use of appropriate technology.

Technology is very essential for collaboration in terms of shared infrastructure, on-demand access, and mechanisms for controlling community privileges [12].

A potential problem that arises is the scaling problem. For instance, the mechanisms that work well for a small amount of participants may not have the same effect for a large amount of participants. This could occur due to issues like trust, shared vocabulary and other implicit knowledge breakdown [11].

A way to solve the scaling problem would be to create infrastructures that grant clients access to associate assertions with data and services. These assertions can be determined as trusted materials on the basis of digital signatures. The consumers can then make their own decisions with regards to quality, authenticity, and accuracy.

### 4.4. Automation of Problem-Solving Protocols

Problem- solving protocols for any science query or computing issues involves the planning and conduction of experiments, collection and analysis of data, derivation of models from data, simulations to explore the implications of models, deduction of new hypotheses from data, and planning new experiments. These steps however, are bound to have some complexity in them which increases over time, making each step applicable for automation [12].

Automated protocols are carried out by computers without human intervention and they operate faster than manual protocols. Therefore, there is a need to document every operation that is being carried out. The documentation is necessary so as to be able to comprehend the protocol; carried out by the computer. These automated protocols are not simple to understand at once, hence why if documented at each step; man would be able to have an idea as to what exactly is going on during the automated problem-solving protocol.

## 5. ENTERPRISE INFORMATION SYSTEM

This section is going to explain what an Enterprise Information System is and would relate Human Computer Symbiosis as to being an Enterprise Information System.

Enterprise Information Systems are very essential in modern organizations and industries because of their ability to develop production and business values, and also set high competitive advantages for such organizations or industries.

They are also known for providing benefits and potential opportunities for the organizations that make use of them [4].

Some key characteristics of Enterprise Information Systems are [15]:

- They are used to integrate information across an organization and often with suppliers and customers.
- They often move from decentralized information systems towards centralized services.
- They are in most cases purchased from a single supplier.
- They are usually an off-the-shelf purchase that can be designed to the requirement desired.

It can be said that a major reason why organizations have Enterprise Information Systems is because of the integrated processes it possesses.

These processes are as follows:
- High speed, efficiency and low cost
- Improved customer service
- Improvement of supplier and customer systems
- Reduction of maintenance costs.
- Reduced complexity of systems
- Straightforward when dealing with just one supplier.

Other key reasons as to why Enterprise Information Systems would be essential include **Information sharing**; better decision making and encouragement of innovation, **Competitive issues**; being on the same level, if not higher, with rival organizations, and opportunities for strategic actions, and **Fragmented data**; being an island full of information [15].

The question now is, "How does Human Computer Symbiosis relate to being an Enterprise Information System".

Human Computer Symbiosis is about integrating the positive characteristics of man and machine in other to perform given tasks in real time. Relating to an Enterprise Information System, they are similar in terms of being an island full of information. Man already has knowledge from past experience or past works being done and computer has a fast processing speed which when integrated successfully with man's knowledge would be able to generate more data needed for the organization, and also aid the organization in terms of production and if possible solving some complex problems the organization might be facing. A successful integration would also help in reducing complexity of systems in the organization.

6. **COMPARISON WITH OTHER INFORMATION SYSTEMS**

This paper has established what an Enterprise information system is all about and has been able to relate Human Computer Symbiosis as an information system. This section would now compare Human Computer Symbiosis with other related information systems.

The other information systems that would be compared with Human Computer Symbiosis are Enterprise Resource Planning and Customer Relationship Management.

An Enterprise Resource Planning (ERP) System is an integrated computer system used to manage internal resources such as tangible assets, financial resources, materials and human resources. The ERP system generates relevant data to all areas in an organization. These distributed data is necessary for various processes such as decisions and data-flow. The ERP system integrates all processes and function in an organization onto a single computer system that could perform all the various processes and functions' needs. It replaces the old standalone computer systems in various departments with a single unified software program [19]. In other words, the ERP system becomes the central information repository in an organization.

The benefits of an ERP system are:
- It synchronizes changes between multiple systems.
- It grants access to the control of business processes.
- It provides real-time view of the enterprise.
- It centralizes permissions and security into a single structure.
- It reduces production lead-time and delivery time.
- It assures information can be shared across all functions and all levels of management to aid the running and managing of the organization.
- The software has a single point of management that handles integration.
- The uniformity or standardization of hardware, software, and data makes room for easier maintenance.

Customer Relationship Management (CRM) systems generates both the data and information on customers in different areas of the organization. The data sources are integrated and shared in order to provide value to the customer. CRM systems provide customer-facing employees with a single, complete view of every customer at each touch point and across all channels. It is a cross-functional enterprise system that integrates many of the customer-serving processes. CRM systems also provide the customer with a single, complete view of the organization and its extended channels.

Benefits of the CRM system are as follows:
- It identifies and targets the best possible customers.
- Real-time customization and personalization of products and services.
- It tracks when and how customers make contact with the organization.
- It provides a consistent customer experience.
- It provides superior service and support across all customer contact points.

The ERP system and the CRM system are in a way similar to each other in terms of generating data and the sharing of information across the organization.

Comparing Human Computer Symbiosis to the ERP and CRM systems, one can conclude that they have similar characteristics. Such characteristics would include the processing of data in real time and the reduction of complexity.

### 6.1. Example of a Potential Problem and Possible Solution

From the comparison of Human computer Symbiosis with other information system, a problem is likely to arise, especially in an organization.

The sharing of information in an organization could end up being a problem rather than a solution. The real question here would be what information is being shared across the organization. Is it sensitive information that is being shared or just basic information? If sensitive information like passwords is shared across the organization, the security is most likely to be comprised, and also the privacy of those who own the passwords, and they are most likely to be that of the customers.

A solution to this problem could be decentralization. By decentralizing the information, there is a limit to who has what information in the organization. On other words, no one would know the whole details of such sensitive information.

Another issue that could arise in Human Computer Symbiosis is that if the gap between human's ability to process information and that of computer's ability is bridged, certain problems such as the NP-Complete would be easily solved. This is an issue because these problems are based on the limitations of computer not being able to see the whole problem at once and humans not being able to evaluate the problem as quickly. If bridged, a lot of security measures used in top companies and banks would be easy to access because these security measures are based on the fact that these problems are unsolvable.

## II. CONCLUSION

Human Computer Symbiosis is a broad topic which relates to various aspects of computing such as cyber-physical-socio spaces and enterprise information system.

It has its advantages which can be very beneficial to man, and it also comes with certain risks which could be harmful.

The focus on this paper has been mainly on the beneficial aspect of human computer symbiosis, stating how the successful integration of human intellect and the computer's processing speed can make the human brain process data better than anything ever known to man.

This paper also looked at the disadvantages of human computer symbiosis bridging the gap between man and computer's ability to process information which could leave many organizations' security very vulnerable.

## III. ACKNOWLEDGEMENT